\documentclass[10pt,showpacs,showkeys,preprintnumbers]{article}
\usepackage{mathrsfs}
\usepackage{amsmath,amssymb,amsfonts}
\usepackage{units}
\usepackage{bbold}
\usepackage{graphicx}
\usepackage{dcolumn}
\usepackage{bm,bbm,mathtools,esvect}
\usepackage{slashed}
\usepackage{esvect}
\usepackage{fullpage}
\usepackage[utf8]{inputenc}
\title{\bf A complementary covariant approach to gravito-electromagnetism}
\author{SERGIO GIARDINO\footnote{\tt sergio.giardino@ufrgs.br}\\
{\small\it Departamento de Matem\'atica Pura e Aplicada}\\ 
{\small\it Universidade Federal do Rio Grande do Sul (UFRGS)}\\
{\small\it Brazil} }

\begin{document}
\date{}
\maketitle

\begin{abstract}
\noindent From a previous paper where we proposed a description of general relativity within the gravito-electromagnetic limit, we propose  an alternative modified gravitational theory.  As in the former version, we  analyze the vector and tensor equations of motion, the gravitational continuity equation, the  conservation of the energy,  the energy-momentum tensor, the field tensor, and the constraints concerning these fields. The Lagrangian formulation is also exhibited as an unified and simple formulation that will be useful for future investigation.

\noindent{\bf keywords:} classical general relativity; fundamental problems and general formalism; modified theories of gravity.

\end{abstract}

\section{ Gravito-electromagnetism}

Is general relativity (GR) a final theory? or it will be 	superseded by another theory in the future? We expect that it will survive while their explanatory power is strong enough to describe the available experimental data. However, even if their explanatory power was not strong enough to understand every known phenomenon, we would keep it while there were not an alternative theory.  At the present time, general relativity is a very successful gravitational theory, but we also know that there are several open questions about it, particularly related to their quantization and to to their cosmological applications. Furthermore, we do not know whether GR is suitable to solve these open questions, or whether a different theory is needed. In this situation, it can be interesting to modify the old theory in order to explain singular data effectively or to introduce a different conceptual idea \cite{Corda:2009re}. From a theoretical point of view, it is interesting to know what kind of modification can possibly be done to a theory and keep their mathematical and physical consistency. 

In this article, we propose an exercise concerning a recently published gravitation theory that modifies GR  within the gravito-electromagnetic precision order \cite{Giardino:2018ffd}. In summary, we will analyze an alternative to this previously  proposed modified theory in order to  exhaust the possible alterations that are coherent to the original idea.
Before considering the new formulation, we give a brief explanation of the origin of gravito-electromagnetism (GEM)
using the Chapter 3 of  \cite{Ohanian:1995uu}. As the name announces, in this theory an analogy between GR and electromagnetism is established. GEM comes fom the weak field approximation to GR, where the $g_{\mu\nu}$ metric tensor  is

\begin{equation}\label{g001}
g^{\mu\nu}=\eta^{\mu\nu}+\kappa h^{\mu\nu},\qquad\qquad\qquad \kappa=\frac{\sqrt{16\pi G}}{c^2}
\end{equation}
is a constant in cgs units  and $\,h^{\mu\nu}\,$ represents the perturbation of the  $\,\eta^{\mu\nu}\,$  flat space tensor, whose components are respectively $\,\eta^{00}=1,\,$ and $\,\eta^{ii}=-1\,$ for $\,i,\,j=\{1,\,2,\,3\}$. Within he weak field limit,  $\,|\kappa h_{\mu\nu}|\ll 1,\,$ the physical law
\begin{equation}\label{g002}
\frac{d\bm v}{dt}=\bm g+\bm{v\times b},
\end{equation}
describes the motion of massive particle of velocity $\,\bm v\,$ in a gravitational
 field $\,\bm g,\,$ and whose gravito-magnetic field is $\,\bm b.\,$ The boldface characters denote vector quantities in a time-like surface, and the vector product satisfies the usual definition. In terms of components, we have
\begin{equation}
\big(\bm{v\times b}\big)_i=\epsilon_{ijk}v_jb_k,
\end{equation} 
where $\epsilon_{ijk}$ is the Levi-Civit\`a anti-symmetric symbol. The similarity between (\ref{g002}) and the electro-dynamical Lorentz force is evident. In terms of the perturbation of the metric tensor, the physical law reads
\begin{equation}\label{g003}
g_i=-\frac{\kappa}{2}\frac{\partial h_{00}}{\partial x^i}\qquad\textrm{and}\qquad 
b_i=-\kappa\left(\frac{\partial h_{0k}}{\partial x^j}-\frac{\partial h_{0j}}{\partial x^k}\right),
\end{equation}
where  the space-like components of the space-time index are $i,\,j$ and $k$. Hence, one can establish an analogy between the covariant electrodynamics and the field vectors in a tensor formula, so that
\begin{equation}\label{g0030}
f_{0i}=g_i\qquad\textrm{and}\qquad f_{ij}=-\frac{1}{2}\epsilon_{ijk}b_k.
\end{equation}
However,  the quantities of (\ref{g003}) were obtained from on the zeroth component of $h_{\mu\nu}$, and hence the   $f_{\mu\nu}$ tensor obtained from (\ref{g0030}) are in fact the zeroth component of a third rank tensor. Therefore, the analogy between $f^{\mu\nu}$ and the Faraday tensor $F^{\mu\nu}$ of electrodynamics is imperfect because the $f^{\mu\nu}$ tensor is not covariant in the same way that $F^{\mu\nu}$. There are various proposals to determine this third rank gravitational tensor, and we quote \cite{Campbell:1970ww,Campbell:1973grw,Campbell:1976ltw,Campbell:1977jf,Ramos:2010zza} and references therein. This fact turns the GEM research even more exciting, because it indicates a way to research a more general tensor theory of gravity, where additional space-time indexes would be necessary to the tensor quantities.
Independently of this feature, the conceptual idea of GEM evidences the parallel between  electromagnetism and gravitation, and various ideas to  implement the gravito-electromagnetic approximation were elaborated, and a list of references of them can be found in \cite{Giardino:2018ffd, Vancea:2021gkf,Behera:2018prr,Behera:2020wva}.

In   \cite{Giardino:2018ffd}, the	 gravitational field $\,\bm g\,$ was decomposed as a sum of two auxiliary fields, the gravito-electric field $\, \bm g_E\,$ and the gravito-magnetic field $\,\bm g_B,\,$ where
\begin{equation}
\bm g=\bm g_E + \bm g_B\qquad\mbox{constrained with}\qquad \bm g_E \bm{\cdot g}_B=0.
\end{equation}
This decomposition is not usual in gravito-electromagnetism (GEM), and the field equations are also different from the previous formulations. This discussion is already been done in the previous article. However, the previous article does not exhaust the possible formulations, and this paper intends to fill this blank. However, we shall see that this task in not a bureaucratic one. The formulations have a diverse physical content, and the second formulation is necessary for the theoretical comprehension, and for  future applications as well.

\section{ Modified Newtonian gravitation \label{I}}

Modified theories of Newton's gravitation are not a novelty, and we mention 
\cite{Milgrom:1983ca,Fischbach:1999bc,Milgrom:2014usa} as a recent conjectures  of such kind.
 In our proposal, the field equations are such as
\begin{equation}\label{g02}
\bm{\nabla\cdot g}=-\,4\pi\rho\qquad\qquad\mbox{and}\qquad\qquad\bm{\nabla\times g}=\frac{4\pi}{c}\bm{p}-\frac{1}{c}\frac{\partial\bm{g}}{\partial t}.
\end{equation}
where $\,\bm g\,$ is the gravity field vector,    $\rho$ is the density of mass, $\,\bm p=\rho\bm v\,$ is the matter flux density vector.
Accordingly, the gravity force $\bm F$   acts over a particle of  mass $m$ according to the physical law
\begin{equation}\label{g03}
\bm F=m\bm g-\frac{1}{c}\bm{p\times g}.
\end{equation}
Equations (\ref{g02}) are identical to that proposed in \cite{Giardino:2018ffd}, while (\ref{g03}) has a single difference in the flipped signal of the second term. The ultimate proposal of the present article is to determine the differences concerning this single difference.
Additionally, we  will confirm that the gravitational field given by (\ref{g02}) has a physical content comparable to that achieved after the truncation of  Eintein's field equations. We remember that truncation of Einstein's equations generate the Newtonian theory at its first approximation, while higher order terms produce (\ref{g002}-\ref{g003}), and this prevision of GR will be recovered from (\ref{g02}) using a covariant scheme. As first consequence, the continuity equation and the conservation of the mass is obtained from (\ref{g02}) 
\begin{equation}\label{g06}
\frac{\partial\rho}{\partial t}+\bm{\nabla\cdot p}=0.
\end{equation}
The energy balance is given by
\begin{equation}\label{g07}
\frac{1}{8\pi c}\frac{\partial |\bm g|^2}{\partial t}+\frac{1}{4\pi}\bm{g\cdot\nabla\times g}=\frac{1}{c}\bm{g\cdot p},
\end{equation}
and  equations (\ref{g06}-\ref{g07}) are identical to that obtained in \cite{Giardino:2018ffd}. We observe the self-interacting terms 
$\,\bm{g\cdot\nabla\times g}\,$, and $\,\bm{g\cdot p}\,$ and a conservative gravitational field is obtained if 
\begin{equation}
\bm{g\cdot\nabla\times g}=0.
\end{equation}
Equation (\ref{g07}) is the gravitational equivalent of the Poynting theorem, but a gravitational Poynting vector cannot be obtained. It is interesting to note that every contribution to the energy balance comes from self-interaction. 
Differently from \cite{Giardino:2018ffd}, an expression for the conservation of the linear
momentum is not possible in this formulation.   Therfore the field equations, the gravitational force law,  the continuity equation and  
the energy balance encompass all the results that one can obtain from this model. In the following section, the tensor approach will illuminate this physical model from a different standpoint.

\section{ The gravitational field in the tensor formalism \label{P}}
In this section, we will observe many differences between the model of (\ref{g02}-\ref{g03}) and the previous article.
Let us then intoduce the gravitational field tensor 
\begin{equation}\label{p03}
C_{\mu\nu}\,=\,\left[
\begin{array}{llll}
0   & -g_1       & -g_2       & -g_3\\
g_1 & \;\;\;0    & -g_3  & \;\;\,g_2\\
g_2 & \;\;\,g_3       & \;\;\,0    & -g_1 \\
g_3 & -g_2  & \;\;\,g_1       & \;\;\,0
\end{array}
\right]\qquad\mbox{where}\qquad\bm g=\big(g_1,\,g_2,\,g_3\big).
\end{equation}
The Minkowskian indices are $\mu$ and $\nu$, and the metric tensor $\eta^{\mu\nu},$ is such that $\,\eta^{00}=1,\,$ $\eta^{ii}=-1\,$
and $i,\,j=\{1,\,2,\,3\}$. Accordingly,
\begin{equation}
C_{i0}=g_i,\qquad\qquad C_{ij}=-\epsilon_{ijk} g_k,
\end{equation}
where $\epsilon_{ijk}$ is the Levi-Civit\`a anti-symmetric symbol. Using the field tensor, the field equations (\ref{g02})
become
\begin{equation}\label{p04}
\partial_\nu C^{\nu\mu}=\frac{4\pi}{c}p^\mu,\qquad\mbox{where}\qquad p^\mu=\big(c\rho,\,\,\bm p\big).
\end{equation}
We also define the contravariant momentum density $4-$vector  $\,p^\mu\,$, that can also be called the matter current $4-$vector, and
the contravariant coordinate $4-$vector $\,x^\mu=\big(ct,\,\bm x\big)\,$.  Using this formalism, the continuity equation (\ref{g06}) reads
\begin{equation}\label{p05}
\partial_\mu p^\mu=0.
\end{equation}
The gravitational force can as well be obtained in the  covariant expression,
\begin{equation}\label{p08}
\frac{d p^\mu}{dt}=\frac{1}{c}C^{\nu\mu}p_\nu.
\end{equation}
The spacelike $\mu=i$ components of (\ref{p08}) furnish the gravity force, and the timelike $\mu=0$ component reveals that the energy density
of the model obeys
\begin{equation}\label{p09}
c^2\frac{d\rho}{dt}=\bm{g\cdot p},
\end{equation}
On the other hand, using the anti-symmetric feature of the field tensor, we obtain 
\begin{equation}\label{p10}
p^\mu\frac{d p_\mu}{dt}=\frac{1}{c}C_{\mu\nu}p^\mu p^\nu=0\qquad\mbox{thus}\qquad \frac{d}{dt} \big(p^\mu p_\mu\big)=0,
\end{equation}
and therefore $p_\mu p^\mu$ reveals to be a constant associated to the rest energy density $E$. Therefore, the 
four-momentum vector (\ref{p04}) can be interpreted relativistically, so that
\begin{equation}\label{p11}
p_\mu p^\mu\,=\,\rho^2c^2-\bm{p\cdot p}=\frac{E^2}{c^2}.
\end{equation}
In order to obtain the energy-momentum tensor of this self-interacting gravitational theory, we define the $\,\tau^{\mu\nu}\,$ symmetric tensor as
\begin{equation}\label{p14}
\tau_{\mu\nu}=\tau^{\mu\nu}=\left\{
\begin{array}{l}
1\qquad\mbox{if}\qquad \mu=\nu,\\
0\qquad\mbox{if}\qquad\mu\neq\nu.
\end{array}
\right.
\end{equation}
which has been introduced in \cite{Giardino:2018ffd} and that satisfies $\,\eta^{\mu\nu}=\tau^{\mu\kappa}\tau_\kappa^{\;\;\nu}.\,$ 
The equations of motion (\ref{p04}) therefore become
\begin{equation}\label{p15}
\partial_\lambda\Big( \tau^\lambda_{\;\;\mu}C_{\nu\kappa}+\tau^\lambda_{\;\;\nu}C_{\kappa\mu}+\tau^\lambda_{\;\;\kappa}C_{\mu\nu}\Big)=
\frac{4\pi}{c}\epsilon_{\mu\nu\kappa\lambda} p_\sigma\tau^{\lambda\sigma},
\end{equation}
where the anti-symmetric Levi-Civit\`a symbol is $\,\epsilon_{\mu\nu\kappa\lambda}.\,$ Consequently, using
(\ref{p04}) and (\ref{p08}), we get
\begin{equation}\label{p16}
\frac{d p_\mu}{dt}=\frac{1}{4\pi}C_{\mu\nu}\partial_\kappa C^{\nu\kappa}.
\end{equation}
Additionally, combining (\ref{p15}-\ref{p16}) produces an equation satisfied by the $T_{\mu\nu}$ is the energy-momentum tensor,
\begin{equation}\label{p17}
\frac{dp^\mu}{dt}=\partial_\kappa T^{\kappa\mu}+I^\mu+S^\mu,
\end{equation}
where $I_\mu$ gives the self-interaction and  $\,S_\mu\,$ represents the source. Explicitly
\begin{equation}\label{p18}
T_{\mu\nu}=\frac{1}{4\pi}\left(C_{\eta\mu}C^{\;\;\eta}_\nu+\frac{1}{4}\tau_{\mu\nu}\tau^\eta_\kappa C_{\eta\lambda}C^{\kappa\lambda}\right),
\qquad I_\mu=\frac{C^{\nu\kappa}}{8\pi}\Big(\tau^\lambda_\nu\tau_\kappa^\eta\partial_\eta C_{\mu\lambda}+\partial_\nu C_{\mu\kappa}\Big),
\qquad
S_\mu=\frac{\epsilon_{\mu\nu\kappa\lambda}}{2c}p_\sigma C^{\kappa\eta}\tau^\nu_\eta\tau^{\lambda\sigma}.
\end{equation}
At this moment, we point out the major difference of  the model presented in this article. This approach is more complicated than the former model \cite{Giardino:2018ffd} where $\,T_{\mu\nu}=I_\mu=0,\,$ and consequently the previous approach is probably more  realistic if we expect that theoretical simplicity and physical reality are twin brothers. Despite this, we further explore the model, and
the energy-momentum tensor and the self-interaction term further simplify to
\begin{equation}\label{p19}
T_{\mu\nu}=\frac{1}{4\pi}\left(C_{\eta\mu}C^{\;\;\eta}_\nu-\frac{1}{2}\tau_{\mu\nu}|\bm g|^2\right)\qquad\textrm{and}\qquad
I_\mu=\frac{1}{4\pi}\Big(\partial_0 C_{\mu i}-\partial_i C_{\mu 0}\Big)C^{0i}.
\end{equation}
Explicitly, the energy-momentum components are
\begin{equation}\label{p20}
T_{00}=\frac{\;|\bm g|^2}{8\pi},\qquad T_{ii}=\frac{1}{8\pi}\Big(|\bm g|^2-4g_i^2\Big),\qquad T_{0i}=0,\qquad T_{ij}=-\frac{g_i g_j}{2\pi},
\end{equation}
which generate the scalar quantities
\begin{equation}\label{p21}
T_{\mu\nu}\tau^{\mu\nu}=0,\qquad T_\mu^{\;\;\mu}=\frac{\;|\bm g|^2}{4\pi}\qquad\mbox{and}\qquad T_{\mu\nu}T^{\mu\nu}=\frac{3|\bm g|^4}{(4\pi)^2}.
\end{equation}
Different from electromagnetism, the gravity energy-momentum tensor is not traceless. This result is in fact expected from general 
relativity, and thus a consistency condition is fulfilled. Furthermore, using the field equations (\ref{g02}), we obtain
\begin{equation}\label{p22}
I^\mu=\frac{1}{4\pi}\left(-\frac{1}{2c}\frac{\partial |\bm g|^2}{\partial t},\;\frac{1}{c}\left(\bm{g\times }\frac{\partial\bm g}{\partial t}\right)_i+
\big(\bm{g\cdot\nabla}\big)g_i\right),\qquad\qquad
S^\mu=\left(\frac{\bm{g\cdot p}}{c},\,-\rho\bm g\right).
\end{equation}
Using (\ref{p20}) and (\ref{p22}) in (\ref{p17}), the energy conservation and the gravitational force components are recovered, and 
the physical consistency of the model is assured. We have shown in this section that the gravitation model that (\ref{g02}-\ref{g03}) 
comprise can be consistently described using a tensor language.  However, such a formulation seems unsatisfactory, particularly because 
the conservation of the energy is not clear in (\ref{g07}). For the sake of clarity, we develop a potential formulation in the next section.

\section{ The gravitational potentials in the tensor formalism \label{PP}}
Introducing  the gravitational scalar potential $\Phi$ and the gravitational vector potential $\bm\Psi$, the gravitational field is proposed to be
\begin{equation}\label{p001}
\bm g=-\bm\nabla\Phi-\frac{1}{c}\frac{\partial\bm\Psi}{\partial t}+\bm{\nabla\times\Psi},
\end{equation}
and the field equations (\ref{g02}) consequently  become
\begin{align}
\nonumber
\nabla^2\Phi\,+\,\frac{1}{c}\frac{\partial}{\partial t}\big(\bm{\nabla\cdot\Psi\big)}&\;=\;4\pi\rho\\
\label{p002}
\nabla^2\bm\Psi-\bm\nabla\big(\bm{\nabla\cdot\Psi}\big)&\;=\;-\frac{4\pi}{c}\bm p\,+\,\frac{1}{c}\frac{\partial}{\partial t}\bm\nabla\Phi\,+\,\frac{1}{c^2}\frac{\partial\bm\Psi}{\partial t^2}.
\end{align}
Nonetheless, we obtain a simpler formulation after defining auxiliary gravito-electric and gravito-magnetic vector fields, respectively
$\,\bm g_E\,$ and $\,\bm g_B.\,$ Therefore,
\begin{equation}\label{p003}
\bm g=\bm g_E+\bm g_B,\qquad\qquad\textrm{where}\qquad\qquad\bm g_E=-\bm\nabla\Phi-\frac{1}{c}\frac{\partial\bm\Psi}{\partial t}\qquad\mbox{and}\qquad\bm g_B=\bm{\nabla\times\Psi}.
\end{equation}
Comparing to the the previous formulation \cite{Giardino:2018ffd} the signals of the third term in (\ref{p001}) and consequently of $\bm g_B$ in (\ref{p003}) are flipped, and the second equation of (\ref{p002}) is simpler than in the previous paper.  
In consequence, using (\ref{p003}) in (\ref{g02}) we obtain the gravitational field equations in potential formulation
\begin{align}
\nonumber&\bm{\nabla\cdot g}_E=-\,4\pi\rho\;\;\;\,\qquad\qquad\qquad \bm{\nabla\cdot g}_B=0\\
\label{p004}&\bm{\nabla\times g}_E=-\frac{1}{c}\frac{\partial\bm g_B}{\partial t}\qquad \qquad\;\;\;\;\;
\bm{\nabla\times g}_B=\frac{4\pi}{c}\bm p-\frac{1}{c}\frac{\partial\bm g_E}{\partial t},
\end{align}
that is similar to previous formulations of GEM \cite{Giardino:2018ffd,Campbell:1970ww,Campbell:1973grw,Campbell:1976ltw,Campbell:1977jf,Ramos:2010zza}, and also similar  to the Maxwell electromagnetic field equations. Defining the gravitational potential second rank tensor 
\begin{equation}\label{p005}
\mathscr{C}^{\mu\nu}\,=\,\tau^\mu_{\;\;\kappa}\tau^\nu_{\;\;\lambda}\Big(\partial^\kappa Q^\lambda-\partial^\lambda Q^\kappa\Big)\qquad
\textrm{where}\qquad Q^\mu=\big(\,\Phi,\,\bm\Psi\,\big)
\end{equation}
is the gravitational potential $4-$vector, we directly have
\begin{equation}\label{p006}
\mathscr C_{i0}=\big(g_E\big)_i\qquad\mbox{and}\qquad \mathscr C_{ij}=-\epsilon_{ijk}\big(g_B\big)_k.
\end{equation}
The potential tensor (\ref{p005}-\ref{p006}) enable us to regain the equations of motion (\ref{p002}) using
\begin{equation}\label{p007}
\partial_\nu\mathscr{C}^{\nu\mu}=\frac{4\pi}{c} p^\mu.
\end{equation}
Equation (\ref{p007})  contains the non-homogeneous components of (\ref{p004}), and the homogeneous terms come from
\begin{equation}\label{p008}
\partial_\lambda\Big( \tau^\lambda_{\;\;\mu}\mathscr{C}_{\nu\kappa}+\tau^\lambda_{\;\;\nu}\mathscr{C}_{\kappa\mu}+\tau^\lambda_{\;\;\kappa}\mathscr{C}_{\mu\nu}\Big)=0.
\end{equation}
Manipulating the $4-$vector momentum density, we consequently have
\begin{equation}\label{p009}
\frac{d p^\mu}{dt}=\frac{1}{c}\,\mathscr{C}^{\nu\mu}p_\nu,
\end{equation}
whose components give
\begin{equation}\label{p010}
\frac{dp^0}{dt}=\frac{1}{c}\bm{p\cdot g}_ E\qquad\qquad\textrm{and}\qquad\qquad\frac{d\bm p}{dt}=\rho\,\bm g_E-\frac{1}{c}\bm{p\times g}_B.
\end{equation}
Analyzing (\ref{p010}) in comparison to (\ref{g03}) and (\ref{p09}), two constraints are emerge, namely
\begin{equation}\label{p011}
 \bm{p\cdot g}_B=0;\qquad c\rho\bm g_B-\bm{p\times g}_E=\bm 0.\qquad\textrm{Likewise,}
 \qquad\bm g_E\bm{\cdot g}_B=0.
\end{equation}
Therefore, the linear momentum $\bm p$, the gravito-electric field $\,\bm g_E\,$ and the gravitational force vector $\,d\bm p/dt\,$ are 
coplanar and the force law (\ref{p011}) conforms perfectly to (\ref{g002}), and the alternated signal  in (\ref{g003})
may be obtained by a redefinition of $\,\bm b\,$. At this moment, we point out the more important drawback of the model. Differently from the previous formulation \cite{Giardino:2018ffd}, we cannot obtain a relation expressing the conservation of momentum in the same fashion as the electromagnetic formulation. This does not means that the momentum is necessarily not conserved, but it may have a more subtle formulation.
We may further explain the conservation of momentum by considering the tensor expression of the force law obtained from (\ref{p007}-\ref{p009}), so that
\begin{equation}\label{p012}
\frac{dp^\mu}{dt}=\partial_\kappa\mathscr{T}^{\kappa\mu}+\mathscr{I}^\mu.
\end{equation}
The energy-momentum tensor is
\begin{equation}\label{p013}
\mathscr{T}_{\mu\nu}=\frac{1}{4\pi}\left(\mathscr{C}_{\lambda\mu}\,\mathscr{C}^{\;\;\lambda}_\nu+\frac{1}{4}\tau_{\mu\nu}\tau^\eta_\kappa\, \mathscr{C}_{\eta\lambda}\mathscr{C}^{\kappa\lambda}\right)=
\frac{1}{4\pi}\left(\mathscr{C}_{\mu\lambda}\,\mathscr{C}^{\;\;\lambda}_\nu-\frac{1}{2}\tau_{\mu\nu}\big|\bm g_B\big|^2\right),
\end{equation}
and the interaction term reads
\begin{equation}\label{014}
\mathscr I^\mu=\frac{1}{4\pi}\left[\,-\frac{1}{2}\partial_0|\bm g_E|^2,
\;\Big(\bm g_E\times\big(\partial_0\bm g_B\big)\Big)_i
+\bm g_E\bm{\cdot\nabla}\big(\bm g_E\big)_i\,\right].
\end{equation}
 The self-interaction term $\,\mathscr{I}^\mu\,$ does not appear in the previous formulation \cite{Giardino:2018ffd}, and this raises up an hypothesis to explain the non-conservative character of the momentum. In electrodynamics, we have to separate the momentum of the particles and the momentum of the fields, and this works well also in \cite{Giardino:2018ffd}. In the present theory, we have the additional contribution of self-interaction of the fields in   (\ref{p012}), and the four-fource cannot be written as a four-divergence, engendering a more general situation here, because such a terms is not present in previous formulations, and the conservation is recovered if $\mathscr{I}^\mu=0$. Maybe we can impose this as a constraint, but this can be considered as a direction for future research, as well as the whole this discussion of the character of momentum in the present theory.

Explicitly written, the components of (\ref{p013}) are
\begin{align}\label{p015}
 &\mathscr T_{00}=\frac{1}{4\pi}\left(\big|\bm g_E\big|^2-\frac{1}{2}\big|\bm g_B\big|^2\right)\;\;\qquad
 \mathscr T_{ii}=\frac{1}{4\pi}\left[\;\frac{1}{2}\big|\bm g_B\big|^2-\big(\bm g_B\big)_i^2-\big(\bm g_E\big)_i^2\;\right]\\
 \nonumber
 &\mathscr T_{0i}=\frac{1}{4\pi}\Big(\bm g_E\times\bm g_B\Big)_i
 \qquad\qquad\qquad\mathscr T_{ij}=-\frac{1}{4\pi}\left[\,\big(\bm g_E\big)_i\big(\bm g_E\big)_j+\big(\bm g_B\big)_i\big(\bm g_B\big)_j\,\right].
\end{align}
Accordingly, we obtain the scalar quantities
\begin{align}\label{p016}
\nonumber\mathscr T_{\mu\nu}\tau^{\mu\nu}&=0,\qquad\qquad
\mathscr T_\mu^{\;\;\mu}=\frac{2\big|\bm g_E\big|^2-\big|\bm g_B\big|^2}{4\pi}\qquad\qquad\mbox{and}\\
\mathscr T_{\mu\nu}\mathscr T^{\mu\nu}&=\frac{2}{(4\pi)^2}\left[\;\frac{|\bm g_B|}{2}^4+|\bm g_E|^4-|\bm g_E|^2|\bm g_B|^2
+\big(\bm g_E\bm{\cdot g}_B\big)^2-\big|\bm g_E\bm{\times g}_B\big|^2\;\right].
\end{align}
By comparing the scalar quantities (\ref{p016}) and (\ref{p20}), the nullity of $\,\mathscr T_{\mu\nu}\tau^{\mu\nu}\,$ and 
$\, T_{\mu\nu}\tau^{\mu\nu}\,$ fits the role played by the null 
$\, T_\mu^{\;\;\mu}\,$ in  electromagnetism. Finally, from (\ref{p012}) we obtain
\begin{equation}\label{p017}
 \frac{dp^0}{dt}=\frac{\partial}{\partial t}\left(\frac{\big|\bm g_E\big|^2-\big|\bm g_B\big|^2}{8\pi}\right)+
 \bm{\nabla\cdot}\left(\frac{\bm g_B\bm{\times g_E}}{4\pi}\right).
\end{equation}
Using (\ref{p010}) we generate the energy conservation law that is directly obtained from the field equations (\ref{p004}) and that does not 
produce additional constraints. Finally,  following a formulation of quantum electrodynamics, we use  $Q^\mu$ from (\ref{p005}), and also $\partial^\mu Q^\mu$,   as the independent variable of the  gravito-electromagnetic Lagrangian density
\begin{equation}\label{p018}
\mathcal{L}=\frac{1}{8\pi}\partial_\mu Q_\nu\mathscr C^{\mu\nu}\,+\frac{1}{c}p_\mu Q^\mu,
\end{equation}
and (\ref{p007}) is immediately obtained from (\ref{p018}).
As a final remark, the field equations (\ref{p004}) can also be obtained using 
\begin{equation}\label{p019}
\bm g=\bm g_E-\bm g_B,\qquad\qquad\textrm{where}\qquad\qquad\bm g_E=-\bm\nabla\Phi+\frac{1}{c}\frac{\partial\bm\Psi}{\partial t}\qquad\mbox{and}\qquad\bm g_B=\bm{\nabla\times\Psi}.
\end{equation}
However, this formulation flips the signal of $\,\bm{p\times g}_B\,$ in (\ref{p010}), and so we conclude that (\ref{p003}) is the most suitable 
choice for the potential. In the next section, we summarize the results of Sections \ref{P} and \ref{PP} into a gravity law that is an
 alternative  to (\ref{g03}).

\section{ The second gravity force law \label{A}}
Let us consider the force law
\begin{equation}\label{a01}
\bm F=\rho\bm g+\frac{1}{c}\bm{p\times g},
\end{equation}
 the field equations
\begin{equation}\label{a02}
\bm{\nabla\cdot g}=-\,4\pi\,\rho,\qquad\bm{\nabla\times g}=-\frac{4\pi}{c}\bm p+\frac{1}{c}\frac{\partial\bm g}{\partial t}.
\end{equation}
and the field tensor 
\begin{equation}\label{a03}
C_{i0}=g_i;\qquad C_{ij}=\epsilon_{ijk}g_k,
\end{equation}
where equations (\ref{p04}-\ref{p09}) hold. On the other hand, (\ref{p15}) changes to
\begin{equation}\label{a04}
\partial_\lambda\Big( \tau^\lambda_{\;\;\mu}C_{\nu\kappa}+\tau^\lambda_{\;\;\nu}C_{\kappa\mu}+\tau^\lambda_{\;\;\kappa}C_{\mu\nu}\Big)=
-\frac{4\pi}{c}\epsilon_{\mu\nu\kappa\lambda} p_\sigma\tau^{\lambda\sigma}.
\end{equation}
The energy-momentum tensor $\,T_{\mu\nu}\,$ is identical to (\ref{p20}), and consequently the scalar
quantities are also identical (\ref{p21}). On the other hand, the source term $\,S^\mu\,$ is identical to that of (\ref{p22}), but
the spacial components of the self-interaction term $\,I^\mu\,$ are slightly different, therefore
\begin{equation}\label{a05}
I^\mu=\frac{1}{4\pi}\left(-\frac{1}{2c}\frac{\partial |\bm g|^2}{\partial t},\;\frac{1}{c}\left(\frac{\partial\bm g}{\partial t}\bm{\times g}\right)_i+
\big(\bm{g\cdot\nabla}\big)g_i\right).
\end{equation}
Hence, the second formulation is also consistent, and the proper physical content demands  experimental investigation 
of  (\ref{g03}) and (\ref{a01}). Additionally, the potential formulation is
\begin{equation}\label{a06}
\bm g=\bm g_E+\bm g_B,\qquad\qquad\textrm{where}\qquad\qquad\bm g_E=-\bm\nabla\Phi+\frac{1}{c}\frac{\partial\bm\Psi}{\partial t}\qquad\mbox{and}\qquad\bm g_B=\bm{\nabla\times\Psi}.
\end{equation}
and finally the field equations are
\begin{align}
\nonumber&\bm{\nabla\cdot g}_E=-\,4\pi\rho\;\;\;\,\qquad\qquad\qquad \bm{\nabla\cdot g}_B=0\\
\label{a07}&\bm{\nabla\times g}_E=\frac{1}{c}\frac{\partial\bm g_B}{\partial t}\qquad \qquad\qquad
\bm{\nabla\times g}_B=-\frac{4\pi}{c}\bm p+\frac{1}{c}\frac{\partial\bm g_E}{\partial t}.
\end{align}
Additionally,
\begin{equation}\label{a08}
\mathscr C_{\mu\nu}=\partial_\lambda\tau_\mu^{\;\;\lambda}Q_\nu-\partial_\lambda\tau_\nu^{\;\;\lambda}Q_\mu,
\end{equation}
leads to,
\begin{equation}\label{a09}
\mathscr C_{i0}=\big(g_E\big)_i,\qquad\qquad \mathscr C_{ij}=\epsilon_{ijk}\big(g_B\big)_k,
\end{equation}
and equations (\ref{p007}-\ref{p008}) are immediately recovered. From (\ref{p009}), we produce
\begin{equation}\label{a10}
\frac{dp^0}{dt}=\frac{1}{c}\bm{p\cdot g}_ E\qquad\qquad\textrm{and}\qquad\qquad\frac{d\bm p}{dt}=\rho\,\bm g_E+\frac{1}{c}\bm{p\times g}_B.
\end{equation}
The constraints are
\begin{equation}\label{a11}
 \bm{p\cdot g}_B=0;\qquad c\rho\bm g_B+\bm{p\times g}_E=\bm 0.\qquad\textrm{Likewise}
 \qquad\bm g_E\bm{\cdot g}_B=0.
\end{equation}
Essentially, both of the formulations are related by the symmetry transformation
\begin{equation}\label{a12}
\bm g_b\to-\bm g_B,\qquad\qquad\textrm{or}\qquad\qquad \bm\Psi\to-\bm\Psi\qquad\qquad\textrm{or}\qquad\qquad Q^\mu\to Q^\nu\tau_\nu^{\;\;\mu}. 
\end{equation}
Thus, under the alternative gravity law the equivalents of (\ref{p012}-\ref{p017}) are immediately obtained using (\ref{a12}), and
the remarkable is the alternate signal in the ``Pointing vector'' of (\ref{p017}), meaning the reversal of
 the momentum flux in each formulation.

\section{ Concluding remarks\label{C}}

We examined several formal questions concerning  gravito-electromagnetism, and proposed two gravity force laws, namely 
(\ref{g03}) and (\ref{a01}), and consistent covariant tensor formulations have been built for both of them. It was also verified that both
of the formulations are related through a symmetry operation. The results complement the former article \cite{Giardino:2018ffd}, where the force law is identical, but the field equations are different different. The results indicate that the energy is conserved in the present formulation, but the momentum is not conserved. Although this seems a negative result, it is in fact a very important piece of information. The force laws (\ref{g03}) and (\ref{a01}) were obtained using a different set of field equations in \cite{Giardino:2018ffd}, and the choices of the present article  introduce the self-interaction terms $\,I^\mu\,$ in (\ref{p17})  and $\,\mathscr{I}^\mu \,$ in (\ref{p012}), and this kind of interaction does not allow the conservation of the momentum.  Only experimental data  concerning the deviation of the Newton law can decide which deviation model generate the correct version of GEM. To the best of our knowledge, the state of the art of the experimental research, namely the Gravity Probe B experiment \cite{Behera:2020wva,Everitt:2015qri}, was unable to pick the most suitable GEM model, and therefore the investigation of the formulations of GEM remains an active field of theoretical research. 

%
%
%
\bibliographystyle{unsrt}

\begin{thebibliography}{10}
\bibitem{Corda:2009re}
{\tt C. Corda}.
\newblock {``Interferometric detection of gravitational waves: the definitive
  test for General Relativity''}.
\newblock {\em Int. J. Mod. Phys.}, {\bf D18}:2275--2282, (2009).

\bibitem{Giardino:2018ffd}
{\tt S. Giardino}.
\newblock {``A novel covariant approach to gravito-electromagnetism''}.
\newblock {\em Braz. J. Phys.}, {\bf 50}(3):372--378, (2020).

\bibitem{Ohanian:1995uu}
{\tt H. Ohanian; R. Ruffini}.
\newblock {``Gravitation and space-time''}.
\newblock Cambridge University Press (2013).

\bibitem{Campbell:1970ww}
{\tt W. B. Campbell; T. Morgan}.
\newblock {``Debye Potentials for the Gravitational Field''}.
\newblock {\em Physica}, {\bf 53}:264--288, (1971).

\bibitem{Campbell:1973grw}
{\tt W. B. Campbell}.
\newblock {``The linear theory of gravitation in the radiation gauge''}.
\newblock {\em Gen. Rel. Grav.}, {\bf 4}:137, (1973).

\bibitem{Campbell:1976ltw}
{\tt W. B. Campbell; T. Morgan}.
\newblock {``Maxwell form of the linear theory of gravitation ''}.
\newblock {\em Am. J. Phys.}, {\bf 44}:356, (1976).

\bibitem{Campbell:1977jf}
{\tt W. B. Campbell; J. Macek; T. A. Morgan}.
\newblock {``Relativistic Time Dependent Multipole Analysis for Scalar,
  Electromagnetic, and Gravitational Fields`''}.
\newblock {\em Phys. Rev.}, {\bf D15}:2156--2164, (1977).

\bibitem{Ramos:2010zza}
{\tt J. Ramos; M. Montigny; F. C. Khanna}.
\newblock {``On a Lagrangian formulation of gravitoelectromagnetism''}.
\newblock {\em Gen. Rel. Grav.}, {\bf 42}:2403--2420, (2010).

\bibitem{Vancea:2021gkf}
{\tt A. Crisan; C. Godinho; I. Vancea}.
\newblock {``Gravitoelectromagnetic knot fields''}.
\newblock {\em Universe}, {\bf 7}(3):46, (2021).

\bibitem{Behera:2018prr}
{\tt H. Behera; N. Barik}.
\newblock {``A New Set of Maxwell-Lorentz Equations and Rediscovery of
  Heaviside-Maxwellian (Vector) Gravity from Quantum Field Theory''}.
\newblock arXiv:1810.04791 [physics.gen-ph] (2018).

\bibitem{Behera:2020wva}
{\tt H. Behera; N. Barik}.
\newblock {``Explanation of Gravity Probe B experimental results using
  Heaviside-Maxwellian (vector) gravity in flat space-time''}.
\newblock 2 2020.

\bibitem{Milgrom:1983ca}
{\tt M. Milgrom}.
\newblock {``A Modification of the Newtonian dynamics as a possible alternative
  to the hidden mass hypothesis''}.
\newblock {\em Astrophys. J.}, {\bf 270}:365--370, (1983).

\bibitem{Fischbach:1999bc}
{\tt E. Fischbach; C. L. Talmadge}.
\newblock {``The search for non-Newtonian gravity''}.
\newblock Springer (1999).

\bibitem{Milgrom:2014usa}
{\tt M. Milgrom}.
\newblock {MOND theory}.
\newblock {\em Can. J. Phys.}, {\bf 93}(2):107--118, 2015.

\bibitem{Everitt:2015qri}
{\tt C. W. F. Everitt; and others}.
\newblock {``The Gravity Probe B test of general relativity''}.
\newblock {\em Class. Quant. Grav.}, {\bf 32}(22):224001, (2015).

\end{thebibliography}
\begin{footnotesize}

\end{footnotesize}
\end{document}